# Sculpting ultrastrong light-matter coupling through spatial matter structuring

Joshua Mornhinweg[1,4,*], Laura Diebel[1], Maike Halbhuber[1], Josef Riepl[1], Erika Cortese[2], Simone De Liberato[2,3], Dominique Bougeard[1], Rupert Huber[1], Christoph Lange[4,*]

[1]Department of Physics, University of Regensburg, 93040 Regensburg, Germany

[2]School of Physics and Astronomy, University of Southampton, Southampton, SO17 1BJ, United Kingdom

[3]IFN - Istituto di Fotonica e Nanotecnologie, CNR, I-20133 Milan, Italy

[4]Department of Physics, TU Dortmund University, 44227 Dortmund, Germany

[*] Corresponding authors

**The central theme of cavity quantum electrodynamics is the coupling of a single optical mode with a single matter excitation, leading to a doublet of cavity polaritons which govern the optical properties of the coupled structure. Especially in the ultrastrong coupling regime, where the ratio of the vacuum Rabi frequency and the quasi-resonant carrier frequency of light, $\Omega_R/\omega_c$, approaches unity, the polariton doublet bridges a large spectral bandwidth $2\Omega_R$, and further interactions with off-resonant light and matter modes may occur. The resulting multi-mode coupling has recently attracted attention owing to the additional degrees of freedom for designing light-matter coupled resonances, despite added complexity. Here, we experimentally implement a novel strategy to sculpt ultrastrong multi-mode coupling by tailoring the spatial overlap of multiple modes of planar metallic THz resonators and the cyclotron resonances of Landau-quantized two-dimensional electrons, on subwavelength scales. We show that similarly to the selection rules of classical optics, this allows us to suppress or enhance certain coupling pathways and to control the number of light-matter coupled modes, their octave-spanning frequency spectra, and their response to magnetic tuning. This offers novel pathways for controlling dissipation, tailoring quantum light sources, nonlinearities, correlations as well as entanglement in quantum information processing.**

# 1 Introduction

Large light-matter coupling strengths have opened up entirely new perspectives for cavity quantum electrodynamics (c-QED): In ultrastrongly coupled structures, exciting quantum phenomena such as the vacuum Bloch-Siegert shift [1], cavity quantum chemistry [2, 3], cavity-controlled electronic transport [4, 5] or the creation of photon-bound excitons [6] have been observed and applications in nanophotonics including novel light sources [7–9], single-photon manipulation [10] and squeezed quantum states of light [11] are being explored. Harnessing such effects requires control of the polariton frequencies, line shapes and oscillator strengths. These properties ultimately result from the underlying exotic ground state of ultrastrong coupling that is characterized by a non-vanishing expectation value of vacuum excitations as well as correlations and entanglement of optical and electronic modes [12, 13]. Landau cavity polaritons combining planar metal cavities with strongly subwavelength field confinement and the large dipole moments of cyclotron resonances (CR) of two-dimensional electron gases in semiconductor quantum wells (QWs) [1, 4, 14–19] have been a highly successful route. The intrinsically large dipole moments of CRs have enabled record coupling strengths of up to $\Omega_R/\omega_c = 3.19$ [19] as well as ultrastrong coupling in few-electron systems using nanogap cavities [18, 20] or ultrastrongly coupled single meta-atoms [21] . Generally, the optical and electronic properties of a light-matter coupled system are defined by the overlap integral of the electric field of the resonator mode and the polarization field of the electronic resonance, each with their specific subwavelength structure [22].

Whereas most investigations have thus far focused on maximising the light-matter coupling strength of a single pair of light-matter coupled modes, or controlling their far-field properties such as polariton frequencies, the large design space accessible by subwavelength control of the shape and overlap of multiple modes has been largely neglected. First explorations include the very-strong coupling regime [6, 23–30] or nanophotonic polariton control in equilibrium [31]. The large frequency range of $2\Omega_R \sim \omega_c$ covered by polaritons of ultrastrongly coupled structures moreover necessitates control of the coupling mechanism over a correspondingly broad spectral range, which can include resonances more than one optical octave above the fundamental cavity mode [32]. Strong nonlinearities and nonlinear polariton mixing was observed in structures featuring multiple polariton modes and corresponding generalized vacuum Rabi frequencies [33]. Moreover, subcycle control of $\Omega_R$ was demonstrated by femtosecond reshaping and switch-off of cavity modes [34], predicted to release the virtual photon population of the exotic vacuum ground state of ultrastrong coupling [34–37]. These examples underscore the importance of the spatial structure of the coupled fields for controlling ultrastrong coupling and its vacuum ground state.

Here, we experimentally explore this design space by tailoring the spatial mode overlap of multiple light and matter modes by structuring the matter component. In similarity to selection rules of classical optics, this new degree of freedom enables us to select particular coupling pathways: nanostructuring controls the extent to which each mode contributes to light-matter hybridization, boosts or suppresses coupling of certain modes, and designs the spectral shape of multiple polariton resonances within a large frequency range. Such an ability is particularly important for large coupling strengths $\Omega_R/\omega_c$, where coupling to multiple modes over an octave-spanning spectral range becomes inevitable and leads to significant resonant and non-resonant interactions [24–26, 28–30, 38]. As a result, multiple coupling pathways influence the non-trivial structure of the vacuum state of ultrastrong coupling at the same time. We demonstrate this concept for two structures which ultrastrongly couple the light field of planar metallic THz resonators to the CR of Landau-quantized two-dimensional electron gases hosted in semiconductor QWs. The reference sample implements a planar QW and is referred to as *unstructured*, while the QW of the second sample, referred to as *structured*, is laterally etched to control the overlap with the resonator modes. Our analysis of experimental data and theory calculations reveals that even without modifying the cavity, this sub-wavelength structuring of the electronic resonance allows for controlling the frequencies, magnetic field dependencies, light-matter coupling strengths, and the number of light-matter coupled modes, and correspondingly, the exotic vacuum ground state. In the future, this idea may increase the level of flexibility for designing multi-mode polaritonic systems and their light-matter couplings and dispersion over multiple octaves in the deep-strong [17, 19, 34] or very-strong coupling regime [6, 23–30].

## 2 Results and discussion

Our resonators (Fig. 1 a) feature rectangular double gaps in the center that provide a capacitive part and are connected to inductive loops defining the outer rectangular perimeter [17]. Within the spectral range of interest here, this design supports a fundamental mode which is characterized by periodic energy exchange between the capacitive and the inductive elements at a frequency of $\nu_1 = 0.8$ THz (LC mode, Fig. 1 b), and correspondingly, a strong near-field enhancement in the center region, which exhibits a phase difference of $\sim\pi$ relative to the outer parts of the resonator (Fig. 2 b). A second mode (dipolar mode) with a resonance frequency of $\nu_2 = 1.6$ THz (Fig. 1 b) partially bypasses the current path through the capacitance and features a more delocalized near-field enhancement which extends to the outer corners of the structure and lacks the phase difference found for the LC mode (Fig. 2 c). These characteristic spatial field distributions are the key to designing the overlap of resonator modes and electronic polarization by structuring the matter system. In particular, they enable control of coupling pathways by continuously tuning the overlap of the cavity modes from zero to unity (Figs. 2 d-f). Our semiconductor heterostructures consist of 3 QWs of a thickness of 20 nm, each n-doped with an electron density of $\rho_{QW} = 1.25 \times 10^{12}$ cm$^{-2}$, and separated by AlGaAs barriers of a thickness of 25 nm. Structuring of the QWs is performed by first covering the heterostructure with resist which is subsequently patterned by electron-beam lithography to form a temporary mask for lateral wet-etching, resulting in a periodic array of rectangular QW patches each measuring 15 µm by 15 µm (Fig. 2 a, orange region). Subsequently, the resonators are fabricated by electron-beam lithography, metal evaporation and a lift-off process, whereby the field-concentrating central gap region is aligned with the etched patches. For the reference sample, the resonators are fabricated on an unstructured QW heterostructure.

Transmission measurements are performed with linearly polarized, single-cycle THz waveforms generated and detected in ZnTe crystals of a thickness of 0.5 mm and 1 mm, respectively. The corresponding amplitude spectra are obtained by Fourier transform and feature a spectral bandwidth ranging from ~200 GHz up to >2 THz. The samples are kept at cryogenic temperatures in a magnet cryostat which provides a static magnetic bias field $B$ of up to 5.5 T applied perpendicularly to the QW plane, introducing Landau quantization of the electrons. The resulting cyclotron resonances are tuneable in frequency, $\nu_c = eB/2\pi m^*$, where $e$ is the elementary charge and $m^* = 0.07\, m_e$ and $m_e$ are the electron effective and free masses, respectively.

We first investigate the unstructured sample implementing the planar QWs, in which the CR couples to both resonator modes without spatial selectivity. In Fig. 3 a, the spectra are plotted as a function of $\nu_c$ as a color plot and reveal five distinct resonances. The diagonal feature at $\nu = \nu_c$ (dashed red line) originates from the CR in uncoupled areas of the structure between resonators. The remaining four dominant modes originate from the light-matter coupling. The lower (LP$_1$) and upper polariton (UP$_1$) modes associated with the first resonator mode emerge at frequencies of 0 THz and 0.9 THz, respectively, for $\nu_c = 0$ THz, and increase in frequency with opposite curvatures as $\nu_c$ is increased, forming the typical anti-crossing shape. Similarly, the corresponding LP$_2$ mode of the second resonator mode branches off the CR near $\nu = \nu_c \approx 1.2$ THz, further increasing in frequency with a declining slope as $\nu_c$ increases. The associated UP$_2$ starts at $\nu = 1.75$ THz, for $\nu_c = 0$ THz, whereby its initially vanishing slope increases with increasing $\nu_c$.

While coupling to the CR requires a vanishing in-plane momentum of the light field, the cavity furthermore generates non-zero in-plane wave vectors which are discretized in energy, resulting from the confinement of the near-field to the central gap region and its periodic structure given by the resonator array, respectively [19]. These components couple to magnetoplasmon modes, resulting in additional light-matter coupled magnetoplasmon-polaritons, in general. Here, we observe only one additional resonance of sizeable oscillator strength, at approximately 0.3 THz above the UP$_1$ mode. Since this feature does not significantly influence the dependence of polariton formation on the spatial structure of the matter component, we restrict our analysis to the LP and UP modes discussed above.

Next, we measure the transmission of the structured sample (Fig. 4 a) where the QW polarization can couple to the near-field only in the central gap area (cf. Fig. 2 a) and hence, the CR signal observed in Fig. 3 is absent. The transmission spectra reveal a fundamentally different polariton spectrum characterized by three rather than four coupled modes. The resonances associated with UP$_1$ and LP$_2$ for the first structure here merge into a single coupled resonance. At $\nu_c = 0$, this mode has a frequency of 0.92 THz which remains approximately constant up to $\nu_c = 0.7$ THz, from where the frequency increases to the inflection point at $\nu_c = 1.25$ THz. As $\nu_c$ is further increased, the frequency converges towards $\nu = 1.6$ THz. The magnetic tuning curve $\nu(\nu_c)$ of this novel feature resembles an S-shape without anti-crossing as theoretically predicted in ref. [31]. In addition, also the frequency response of the other coupled resonances differs from the situation in the unstructured sample, while the absorption strength of the polariton resonances remains comparable (see Supplementary Material). In more universal terms, the very large coupling strengths of, e.g., the deep-strong coupling regime, inevitably cause all fundamental resonances to influence each other even across an ultrabroadband spectral range. Tailoring the spatial overlap of these modes by near-field structuring allows us to take control of the of individual coupling constants that link pairs of optical and electronic modes of a light-matter coupled structure, in analogy to selection rules of classical optics (Fig. 2 d-f).

In a first approach, we describe our data by a classical electrodynamical formalism employing finite-element frequency-domain (FEFD) simulations [17], which reproduce the transmission spectra and the frequency response of all modes with high accuracy, without free fit parameters (Figs. 3 b and 4 b). These calculations furthermore deliver the spatial profiles of uncoupled (see Figure 2) and coupled modes, enabling the identification of the polariton resonances. To explain the observed differences between both structures and extract the coupling and overlap parameters, we developed a novel quantum mechanical formalism reported in detail elsewhere [31], which extends the established Hopfield model to non-orthogonal modes in order to take the fractional overlap of light and matter modes of our structured sample into account. The bosonic Hamiltonian reads:

$$\hat{H} = \sum_\nu \hbar\omega_\nu \hat{a}_\nu^\dagger \hat{a}_\nu + \sum_\nu \hbar\omega_c \hat{b}_\nu^\dagger \hat{b}_\nu$$
$$+ \sum_\nu \sum_{\mu \leq \nu} \hbar \left[ (\Omega_{R,\nu,\mu} \hat{b}_\mu + \Omega_{R,\nu,\mu}^* \hat{b}_\mu^\dagger)(\hat{a}_\nu^\dagger + \hat{a}_\nu) \right]$$

(1)

$$+ \sum_{\nu,\mu} h_{\nu,\mu}(\hat{a}_\nu^\dagger + \hat{a}_\nu)(\hat{a}_\mu^\dagger + \hat{a}_\mu),$$

where the first two sums describe the bare cavity and matter systems. Here, the bosonic annihilation operator $\hat{a}_\nu$ is associated with the $\nu$-th cavity mode with a frequency $\omega_\nu = 2\pi\nu_\nu$, and $\hat{b}_\mu$ is the collective bosonic matter operator which describes the $\mu$-th degenerate Landau excitation with frequency $\omega_c = 2\pi\nu_c$. The third double sum describes coupling pathways between all light and a matter modes. Here,

$$h_{\nu,\mu} = \sum_{\gamma \leq \nu,\mu} \frac{\Omega_{R,\nu,\gamma}\Omega_{R,\mu,\gamma}}{\omega_c} \quad (2)$$

contains the coupling parameters $\Omega_{R,\nu,\gamma}$, i.e., the vacuum Rabi frequencies quantifying the spatial overlap and thus the mutual interaction between modes $\nu, \gamma$, whereby $\frac{\Omega_{R,\nu,\gamma}}{\omega_c}$ quantifies the associated normalized coupling strength. Moreover, this term accounts for interactions caused by the non-orthogonality of modes in the limited subspace of interaction, i.e., the QW plane [31], as compared to the full three-dimensional space, where orthogonality holds. Note that within a local response approximation, we consider the matter modes as degenerate since they exhibit different real-space wavefunctions, whereas in systems with ultra-narrow features one would be obliged to take the dispersion of the magneto-polaritons into account [30]. The last sum of Eq. 1 is the diamagnetic term that describes the cavity blue shifts for each resonator mode arising from self-interaction. Owing to the bosonic nature of the Hamiltonian, the standard Hopfield diagonalization can be performed.

As shown in greater detail in [31], for two photonic modes we can express the coupling strengths in terms of a single overlap parameter $\eta_{2,1}$ which we calculate based on the electric near-field distribution of the resonator modes (see Supplemental Material) and which assumes values between 0 and 1:

$$\Omega_{R,1,1}, \quad \Omega_{R,2,1} = \tilde{\Omega}_{R,2}\eta_{2,1}, \quad \Omega_{R,2,2} = \tilde{\Omega}_{R,2} \cdot \sqrt{1 - |\eta_{2,1}|^2}. \quad (3)$$

For the matter excitations, we chose a basis for which the photonic mode $\nu = 1$ is always exclusively coupled to the matter mode $\mu = 1$, implying that $\Omega_{R,1,2} = 0$, while the coupling of the second mode, $\widetilde{\Omega}_{R,2}$, is redistributed between both matter modes for a non-zero overlap parameter $\eta_{2,1}$. While higher-order photonic modes are present in our structures, their coupling strength and influence is negligible (see Supplementary Material).

We employ our multi-mode formalism to fit the resonances of the spectra calculated as a function of $\nu_c$ (Figs. 3 b, 4 b, black curves). The coupling parameters $\Omega_{R,1,1}$ and $\widetilde{\Omega}_{R,2}$ serve as fitting parameters. Comparing these spectra for the two configurations enables us to determine the key differences. First, only the fit in Fig. 3 b exhibits the S-shaped magnetic tuning curve. This distinct feature can be understood by comparing the overlap parameter which assumes $\eta_{2,1} = 0.15$ for the unstructured system and $\eta_{2,1} \cong 1$, for the structured sample. Considering that in principle, $0 < \eta_{2,1} < 1$, these non-zero overlap factors show that interactions of the photonic modes are present also in the case of the unstructured sample. Whereas the global orthogonality of the resonator modes rules out interactions, the restriction of the light-matter interaction volume to a finite sub-domain represented either by the full QW plane or by the etched QW patch lifts this restriction, allowing us to tailor the interaction pathways to a large degree. Correspondingly, the two values of $\eta_{2,1}$ lead to strongly different polariton spectra, where the S-shaped spectral signature is a hallmark of nearly full mode overlap, $\eta_{2,1} \approx 1$. Our model shows that this spectral shape indicates a change of character of the light-matter coupling mechanism which manifests in a reduction of the number of participating modes from 4 to 3 as $\eta_{2,1}$ approaches unity. This can be formally seen from Eq. 3, where $\Omega_{R,2,2} = 0$ for $\eta_{2,1} = 1$, i.e., the interaction between the second photon mode and the second electronic mode (orthogonal to and degenerate with the first one), vanishes, and the cross-interaction term proportional to $\Omega_{R,2,1}$ becomes dominant. Therefore, the same electronic mode interacts with both photon modes $\nu = 1$ and $\nu = 2$, which confine the resulting coupled mode in the spectral range enclosed by them, force the formation of an inflection point, and mark the asymptotic frequencies attained for $\nu_c \to 0$ and $\nu_c \to 2$ THz, respectively. The uncoupled electronic mode follows $\nu = \nu_c$ and thus produces a diagonal spectral signature identical to that of the CR in areas between resonator structures. As an additional result, we observe a general reduction of both the polariton splitting as well as the two modes' diamagnetic blue shifts, resulting in reduced normalized coupling strengths of $\frac{\Omega_{R,1,1}}{\omega_1} = 0.28$, $\frac{\Omega_{R,2,2}}{\omega_2} = 6 \times 10^{-3}$, and $\frac{\Omega_{R,2,1}}{\omega_1} = 0.27$, as compared to the unstructured case where $\frac{\Omega_{R,1,1}}{\omega_1} = 0.37$, $\frac{\Omega_{R,2,2}}{\omega_2} = 0.21$, and $\frac{\Omega_{R,2,1}}{\omega_1} = 0.07$. While overall, the coupling strengths decrease when structuring the QWs, the relative weight of the DP mode, indicated by the ratio $\Omega_{R,2,1}/\Omega_{R,1,1}$, increases from 0.19 to 0.96 upon introducing the QW patch. A simulation systematically sweeping the patch size shows how the parameters change from minimum to maximum mode overlap (see Supplementary Material).

**3 Conclusion**

In conclusion, the concept of lateral confinement of the quantum wells adds a previously unexplored parameter space for tailoring ultrastrong light-matter coupling by controlling the spatial mode overlap in a situation of multiple interacting light and matter modes, across several optical octaves. As we have experimentally demonstrated, the character of the resulting polariton modes is strongly controlled by the overlap, including polariton frequencies, magnetic field dependencies, and spatial field distributions which are relevant for nanophotonic applications. Since for deep-strong coupling conditions, coupling of multiple modes over an ultrabroadband range is almost inevitable, future implementations following our approach may benefit from the ability to selectively boost or suppress certain coupling pathways, in similarity to tailoring of selection rules in classical optics. This novel design approach increases the flexibility for sculpting polaritonic structures by exploiting mutual couplings, controlling modal dispersion, or tailoring resonator structures with strongly non-overlapping optical modes.

**Acknowledgments**: The authors thank Dieter Schuh and Imke Gronwald for valuable discussions and technical support.

**Research funding**: We gratefully acknowledge support by the Deutsche Forschungsgemeinschaft (dx.doi.org/10.13039/501100001659) through Project IDs 422 31469 5032-SFB1277 (Subproject A01), grants no. BO 3140/3-2, LA 3307/1-2, and HU 1598/8, as well as by the European Research Council (ERC) (dx.doi.org/10.13039/100010663) through Future and Emerging Technologies (FET) grant no. 737017 (MIR-BOSE) and by the Leverhulme Trust thought the grant RPG-2022-037.

**Author contribution**:

J.M., D.B., R.H. and C.L. conceived the study. L.D., M.H. and D.B. designed, realized, and characterized the semiconductor heterostructures. J.M. and L.D. modelled the metasurfaces and fabricated the samples with support from M.H., J.R., D.B. and C.L.. J.M., M.H., L.D. and J.R. carried out the experiments with support from R.H. and C.L.. The theoretical modelling was carried out by E.C. and S.D.L..S.D.L., D.B., R.H. and C.L. supervised the study. All authors analyzed the data and discussed the results. J.M., E.C., S.D.L., R.H. and C.L. wrote the manuscript with contributions from all authors. All authors have accepted responsibility for the entire content of this manuscript and approved its submission.

**Conflict of interest**: Authors state no conflict of interest.

**Data availability statement**: The datasets generated and/or analyzed during the current study are available from the corresponding author upon reasonable request.

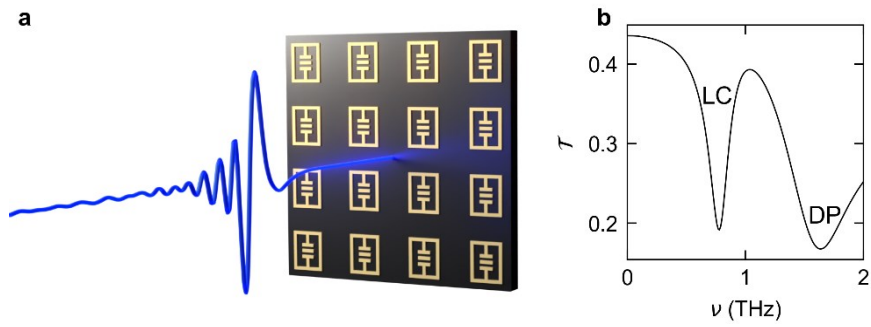

**Figure 1 | Ultrastrongly coupled sample structure. a**, Schematic of the structure consisting of an array of THz resonators (gold shapes), fabricated on top of the GaAs substrate (dark brown). The transmission is probed by broadband, few-cycle THz transients (blue waveform). **b**, Simulated transmission spectrum of the bare resonator structure. LC: fundamental mode with a frequency of $v_1 = 0.8$ THz. DP: higher-order mode with a frequency of $v_2 = 1.6$ THz.

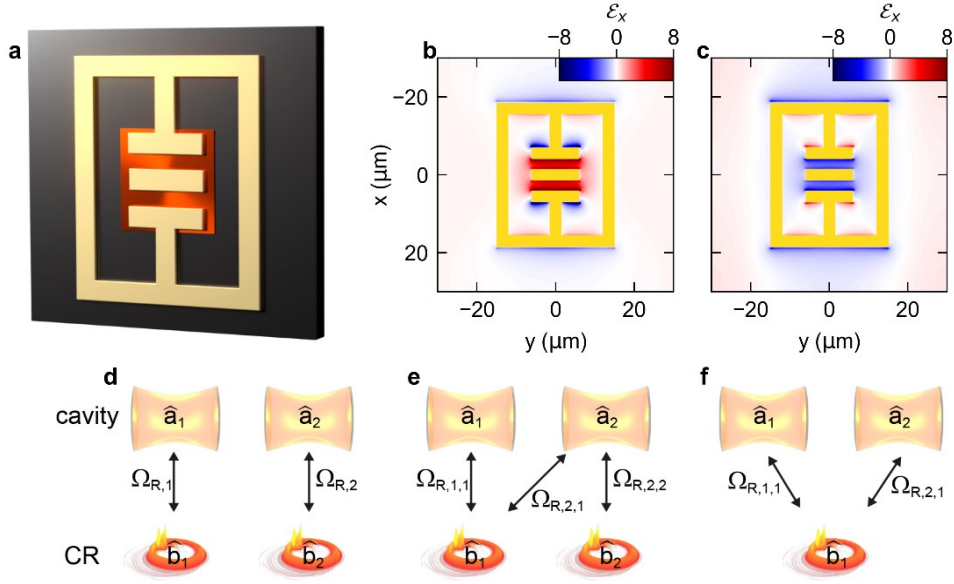

**Figure 2 | Mode-selective ultrastrong coupling by nanostructured matter resonances.**
**a**, Schematic of a single THz resonator (gold shape) fabricated on top of the GaAs substrate (dark brown), and the cyclotron resonances hosted in spatially confined, high-quality GaAs quantum well structures (bright red patch). **b**, Simulated x-polarized near-field distribution of the LC mode of the bare resonator, and **c**, the dipolar mode. **d**, Schematic of the coupling with zero mode overlap, **e**, partial overlap and **f**, full overlap of the 1st (LC, $\nu = 1$) and 2nd (DP, $\nu = 2$) cavity modes.

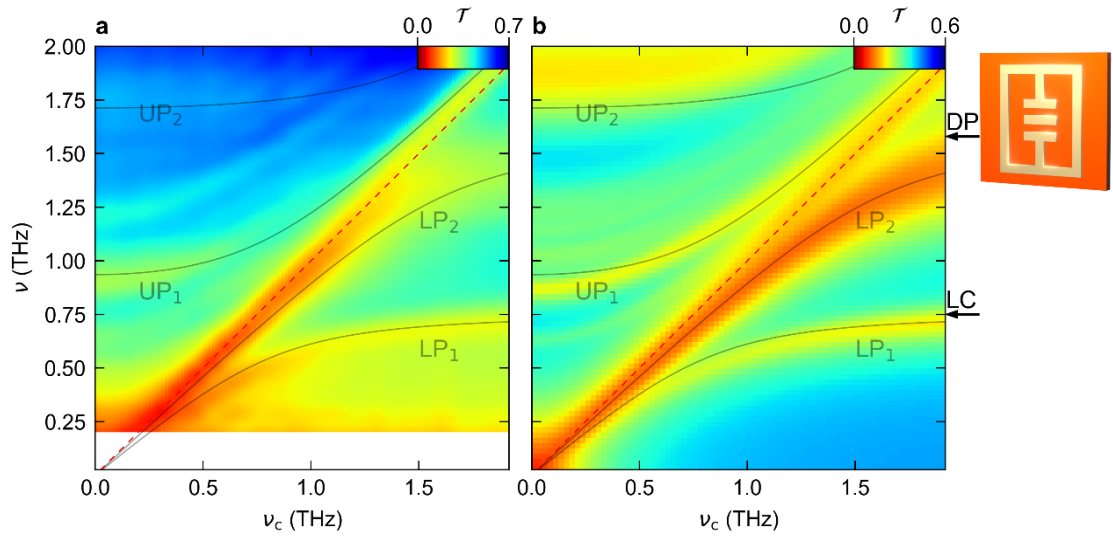

**Figure 3 | Multi-mode coupling with an unstructured QW film. a**, Experimental transmission spectra of the unstructured sample as a function of the cyclotron frequency $\nu_c$. The semi-transparent curves represent the eigenmodes calculated by theoretical fit. **b**, FEFD simulation of the transmission spectra of the structure as a function of the cyclotron frequency $\nu_c$, including the eigenmodes of the theoretical fit in panel **a**. The two arrows indicate the frequencies of the uncoupled LC and DP modes obtained by our fit. Outset: Schematic of the resonator on top of an unstructured QW (bright red area).

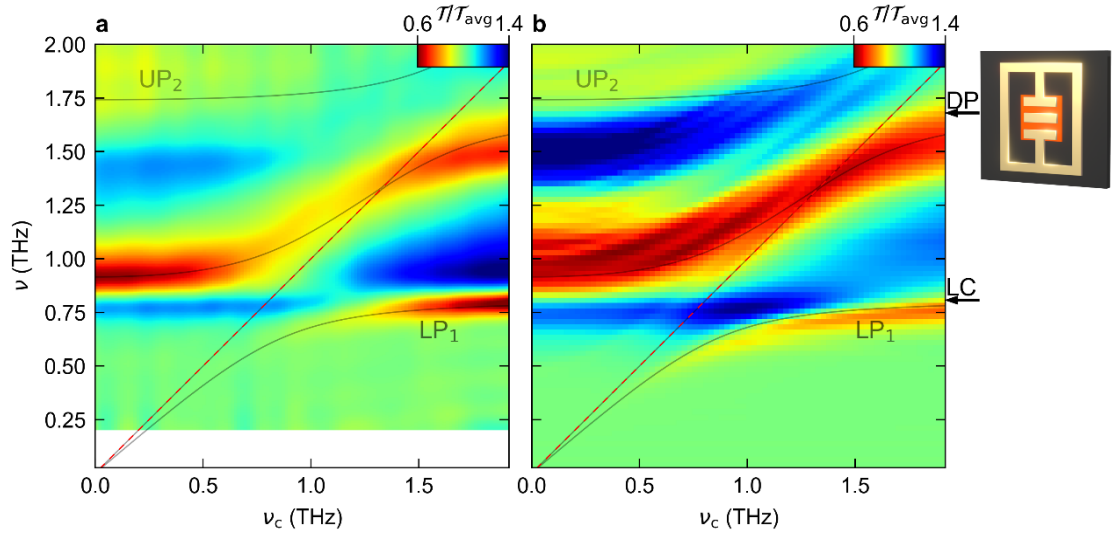

**Figure 4 | Multi-mode coupling with a structured QW film. a**, Experimental transmission spectra of the unstructured sample, each normalized to the average transmission $\mathcal{T}_{avg}$ for each frequency $\nu$, as a function of the cyclotron frequency $\nu_c$. The semi-transparent curves represent the eigenmodes calculated by theoretical fit. **b**, FEFD simulation of the transmission spectra of the structure, each normalized to the average transmission $\mathcal{T}_{avg}$ for each frequency $\nu$ as a function of the cyclotron frequency $\nu_c$, including the eigenmodes of the theoretical fit in panel **a**. The two arrows indicate the frequencies of the uncoupled LC and DP modes obtained by our fit. Outset: Schematic of the resonator on top of a structured QW patch (bright red patch).

# Sculpting ultrastrong light-matter coupling through spatial matter structuring
# Supplementary Material

Joshua Mornhinweg[1,4,*], Laura Diebel[1], Maike Halbhuber[1], Josef Riepl[1], Erika Cortese[2], Simone De Liberato[2,3], Dominique Bougeard[1], Rupert Huber[1], Christoph Lange[4,*]

[1]*Department of Physics, University of Regensburg, 93040 Regensburg, Germany*

[2]*School of Physics and Astronomy, University of Southampton, Southampton, SO17 1BJ, United Kingdom*

[3]*IFN - Istituto di Fotonica e Nanotecnologie, CNR, I-20133 Milan, Italy*

[4]*Department of Physics, TU Dortmund University, 44227 Dortmund, Germany*

[*]*corresponding authors*

## Table of Contents

# 1. Theory of multi-mode coupling

In ref. [31] we developed a theory for the light-matter coupling between a given number of photonic resonator modes and the cyclotron resonances (CRs) of a 2DEG of carrier density $\rho_{QW}$. Whilst the full derivation of the multi-mode Hamiltonian is detailed in [31], we summarize the key points below.

The full vector potential of the electromagnetic field of the resonator can be expressed as a sum of the photonic modes with dimensionless spatial profile $f_\nu(\mathbf{r})$, frequency $\omega_\nu$, and photonic annihilation operator $\hat{a}_\nu$, as

$$\hat{A}(\mathbf{r}) = \sum_\nu \sqrt{\frac{\hbar}{2\,\epsilon_0 \epsilon_r \omega_\nu V_\nu}}\, f_\nu(\mathbf{r})\,(\hat{a}_\nu + \hat{a}_\nu^\dagger), \qquad (1)$$

where $V_\nu$ is the mode volume, and $\epsilon_r$ is the background dielectric constant in the QW plane. Placing a 2DEG below the resonator reduces the dimensionality of the system and breaks the orthogonality of the different photonic modes, displayed by the non-zero integral matrix element

$$F_{\nu,\mu} = \int_S f_\nu^*(z, r_\parallel)\, f_\mu(z, r_\parallel)\, dr_\parallel, \qquad (2)$$

where $r_\parallel$ is the in-plane vector, $z$ is the out-of-plane coordinate of the 2DEG, and $S$ is the in-plane domain, i.e., the QW surface. By remapping the non-orthogonal photonic modes $f_\nu(z, r_\parallel)$ onto an arbitrary basis of orthogonal in-plane functions $\phi_\nu(z, r_\parallel)$ as

$$f_\nu(z, r_\parallel) = \sum_\mu \alpha_{\nu,\mu}\, \phi_\nu(z, r_\parallel), \qquad (3)$$

and, substituting Eq. 3 into the expression of the vector potential, one obtains a Hopfield-like Hamiltonian, which includes cross-interaction terms

$$\hat{H}_{int} = \sum_\nu \sum_{\mu \leq \nu} \hbar\left[(\Omega_{R,\nu,\mu} \hat{b}_\mu + \Omega_{R,\nu,\mu}^* \hat{b}_\mu^\dagger)(\hat{a}_\nu^\dagger + \hat{a}_\nu)\right], \qquad (4)$$

where a set of collective bosonic matter operators $\hat{b}_\nu$ represents in-plane degenerate but orthogonal modes for the electronic excitations.

The cross-interaction between the $\nu^{th}$ photon mode and $\mu^{th}$ electronic mode is described by the vacuum Rabi energies $\Omega_{R,\nu,\mu} \propto \alpha_{\nu,\mu}$, which can be expressed in terms of an overlap parameter defined as $\eta_{\nu,\mu} = \frac{F_{\nu,\mu}}{\sqrt{F_{\nu,\nu} F_{\mu,\mu}}}$. The parameter $\eta_{\nu,\mu}$ attains a value of 1 when the field distribution of the two modes is identical within the domain $S$. This is the case in our structured sample where the domain $S$ is restricted to the central region of the resonator, where both modes overlap and differ only by a normalising factor which is included in an effective mode length $\tilde{V}_\nu = \frac{V_\nu}{F_{\nu,\nu}}$ in z-direction.

In the case of two photonic modes, the coupling strengths are given in full as:

$$\Omega_{R,1,1} = \sqrt{\frac{\omega_c n_{QW} \rho_{QW} e^2}{2 m^* \epsilon_0 \epsilon_r \omega_1 \tilde{V}_1}}, \tag{5}$$

$$\Omega_{R,2,1} = \sqrt{\frac{\omega_c n_{QW} \rho_{QW} e^2}{2 m^* \epsilon_0 \epsilon_r \omega_2 \tilde{V}_2}} \eta_{2,1}, \tag{6}$$

$$\Omega_{R,2,2} = \sqrt{\frac{\omega_c n_{QW} \rho_{QW} e^2}{2 m^* \epsilon_0 \epsilon_r \omega_2 \tilde{V}_2}} \sqrt{1 - |\eta_{2,1}|^2}. \tag{7}$$

## 2. Alternative visualisation of the data

In addition to the color maps shown in Figs. 3 and 4 of the manuscript, we provide the same data as waterfall plots in Figs. S1 and S2. The curves are ordered by the value of the cyclotron resonance frequency, $\nu_c$, in vertically ascending order.

Moreover, since the LP$_1$ resonance is difficult to track in the 2D colormap of Fig. 4a owing to the comparably low oscillator strength, we show the magnetic field dependence of the LP$_1$, normalized to the average transmission $\mathcal{T}_{avg}$ for each frequency $\nu$, in Fig. S3.

For a quantitative comparison of the oscillator strengths of the polariton resonances, we extract the values of the absorption of several of the coupled modes. For the experimental data, the transmission of the LP$_1$ and UP$_1$ modes of the unstructured sample reaches $\mathcal{T}_{LP_1} \approx 0.26$ at $\nu_c = 1.9$ THz, and $\mathcal{T}_{UP_1} \approx 0.32$, at $\nu_c = 0$ THz. In comparison, the absorption of the polariton modes of the structured sample vary more strongly, but reach similar minimal values of $\mathcal{T}_{LP_1} \approx 0.21$ at $\nu_c = 1.9$ THz, and $\mathcal{T}_{S-mode} \approx 0.24$, at $\nu_c = 0$ THz. A similar result is obtained from the FEFD simulations with $\mathcal{T}_{LP_1} \approx 0.22$ and $\mathcal{T}_{UP_1} \approx 0.22$ for the unstructured sample and $\mathcal{T}_{LP_1} \approx 0.21$ and $\mathcal{T}_{S-mode} \approx 0.21$ for the structured one, obtained in each case for the same $\nu_c$ as for the corresponding experimental data.

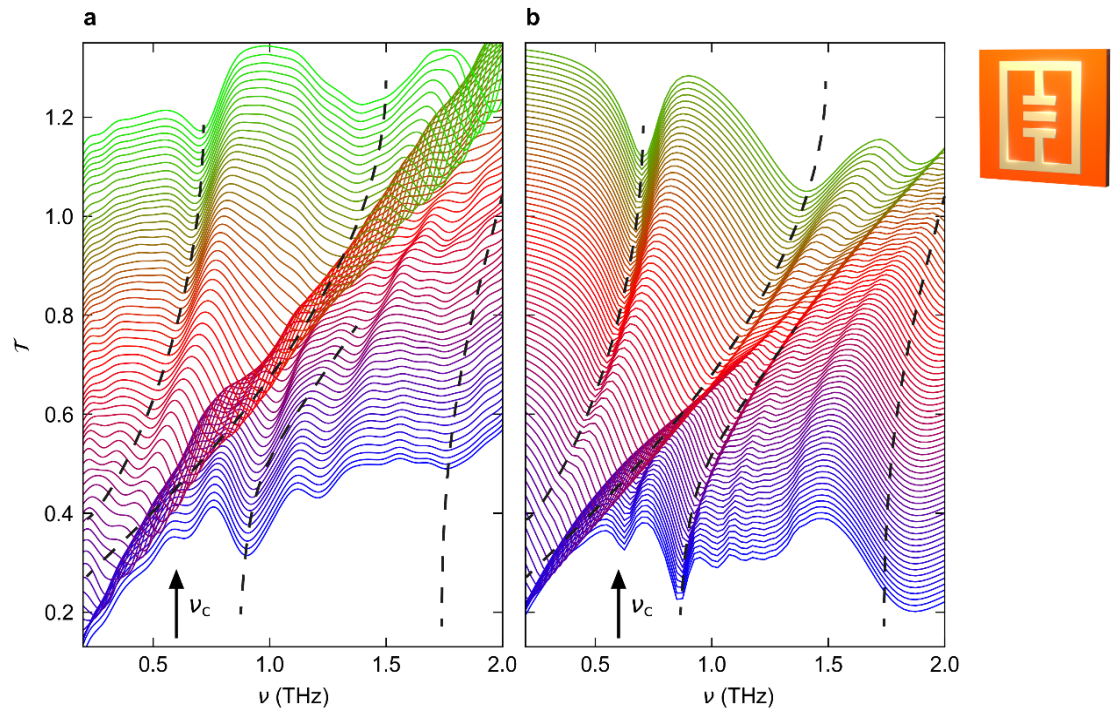

**Figure S1: a**, Experimental transmission spectra of the unstructured sample as a function of the cyclotron frequency, $\nu_c$. The individual spectra are vertically offset for visual clarity. Dashed lines as a guide to the eye. **b**, Corresponding FEFD simulation of the transmission spectra. The individual spectra are vertically offset for visual clarity. Dashed lines as a guide to the eye. Outset: Schematic of the resonator on top of an unstructured QW (bright red area).

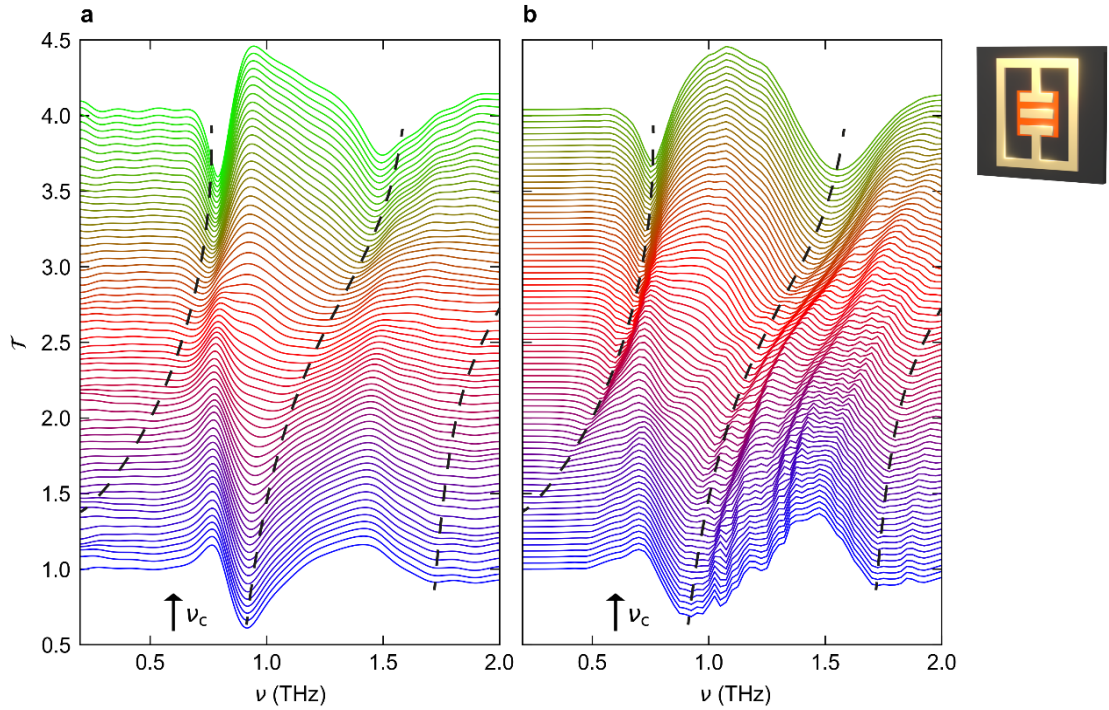

**Figure S2: a**, Experimental transmission spectra of the structured sample, each normalized to the average transmission $\mathcal{T}_{avg}$ for each frequency $\nu$, as a function of the cyclotron frequency $\nu_c$. The individual spectra are vertically offset for visual clarity. Dashed lines as a guide to the eye. **b**, Corresponding FEFD simulation of the transmission spectra. The individual spectra are vertically offset for visual clarity. Dashed lines as a guide to the eye. Outset: Schematic of the resonator on top of a structured QW (bright red area).

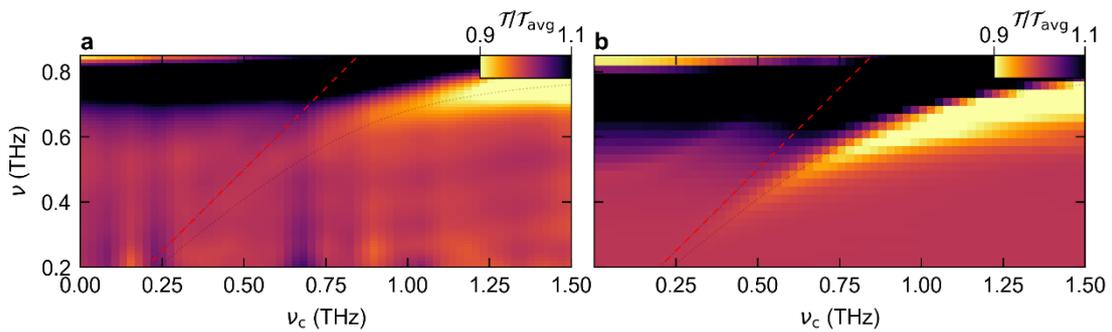

**Figure S3: a**, Experimental transmission spectra of the unstructured sample, each normalized to the average transmission $\mathcal{T}_{avg}$ for each frequency $\nu$, as a function of the cyclotron frequency $\nu_c$. The semi-transparent curves represent the eigenmodes calculated by our theoretical model. **b**, FEFD simulation of the transmission spectra, each normalized to the average transmission $\mathcal{T}_{avg}$ for each frequency $\nu$, as a function of the cyclotron frequency $\nu_c$, including the theoretical eigenmodes of panel **a**.

## 3. FEFD simulations of light-matter coupling

We perform finite-element frequency-domain (FEFD) simulations of the transmission of our coupled structures which solve Maxwell's equations on a discretized version of the sample geometry consisting of the GaAs substrate, the GaAs QW stack, the gold resonator structure, and vacuum. Details of this approach are given in Ref. [17]. We implement the QW response by a gyrotropic dielectric tensor parameterized by the cyclotron resonance frequency and an oscillator strength proportional to the charge carrier density. For the latter, we chose a value of $\rho_{QW} = 1.25 \times 10^{12}$ cm$^{-2}$ for best agreement with the experiment, which is in close vicinity of the nominal chemical doping density of $1 \times 10^{12}$ cm$^{-2}$.

In an additional set of simulations, we investigate the role of higher-order modes. Generally, in a multi-mode light-matter coupled structure, modes within a spectral range comparable to or smaller than their vacuum Rabi frequencies, $\Omega_R$, need to be considered. FEFD simulations for a frequency range of up to $\nu = 6$ THz and up to $\nu_c = 6$ THz (Fig. S4) show that higher-order photonic modes n ≥ 3 exhibit almost no light-matter coupling signatures and thus contribute only negligibly to the coupling mechanism in our structures. We have thus restricted the analysis to the LC and DP modes.

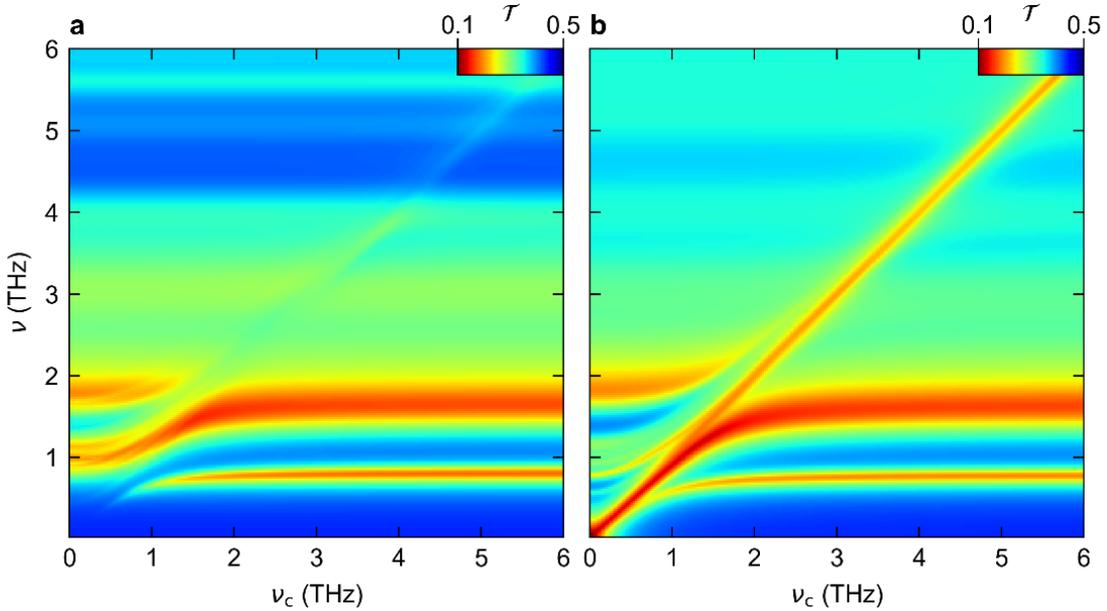

**Figure S4**: Wide-band FEFD simulation of the transmission spectra of **a**, the structured and **b**, the unstructured sample, as a function of the cyclotron resonance frequency, $\nu_c$.

## 4. Dependence of polariton formation as a function of the patch size

In addition to the simulations supporting the two experimentally investigated cases, we provide additional simulations and fits based on our Hamiltonian model (Fig. S5), which systematically show the transition of the relevant coupling parameters between the two extremal cases. The numerical values are given in Table S1. The data show a clear, monotonic progression of the frequencies of the coupled modes as well as the vacuum Rabi frequencies, from the case of the smallest investigated patch size of 15 μm side length to the full, unstructured QW film. Moreover, the merging of the $LP_2$ and $UP_1$ resonances into the S-shaped mode is traced.

Finally, Fig. S6 shows the overlap parameter $\eta_{2,1}$ as a function of the patch size. The data indicate a comparably sharp transition from full overlap, $\eta_{2,1} = 1$, to the limiting value of $\eta_{2,1} = 0.15$ for an unstructured, infinitely extended QW system. The experimentally investigated structures thus represent the extremes of both scenarios.

| Patch side length [μm] | 15 | 20 | 25 | 30 | 35 | Full QW |
|---|---|---|---|---|---|---|
| $\frac{\Omega_{R,1,1}}{\omega_1}$ | 0.28 | 0.37 | 0.45 | 0.43 | 0.37 | 0.37 |
| $\frac{\Omega_{R,2,1}}{\omega_1}$ | 0.27 | 0.31 | 0.38 | 0.27 | 0.26 | 0.07 |
| $\frac{\Omega_{R,2,2}}{\omega_2}$ | 0.006 | 0.005 | 0.01 | 0.03 | 0.08 | 0.21 |
| $\eta_{2,1}$ | 0.999 | 0.9995 | 0.9975 | 0.97 | 0.84 | 0.15 |

**Table S1:** Coupling parameters for the simulations with varying QW patch size.

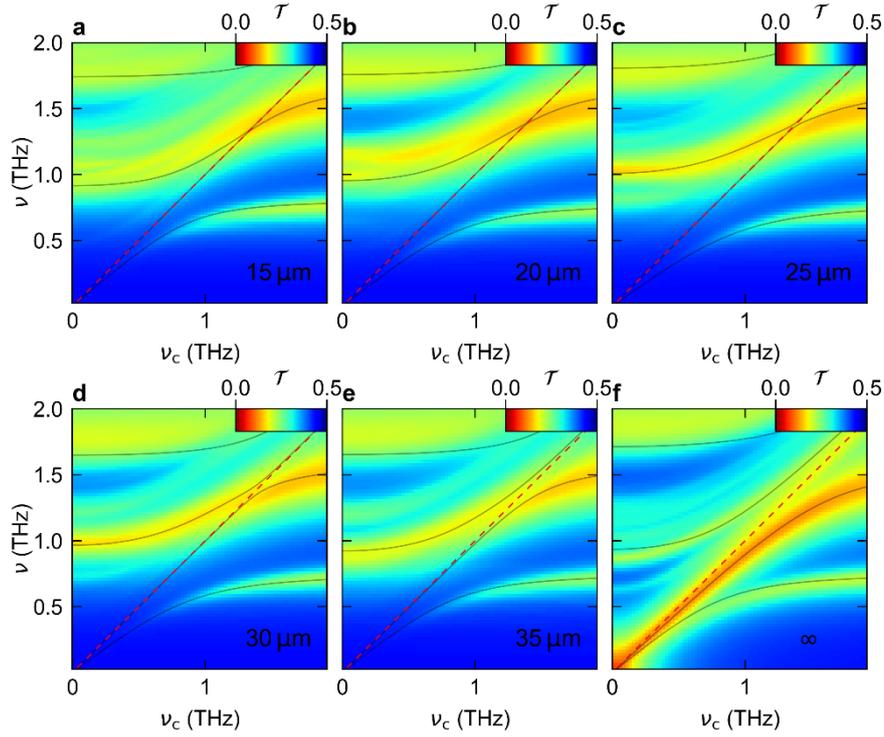

**Figure S5**: FEFD simulation of the transmission spectra of the structures as a function of the cyclotron frequency $\nu_c$ with different patch sizes and including individual fits with the multi-mode Hamiltonian. **a**, structured QW with a side length of 15 μm, **b**, 20 μm, **c**, 25 μm, **d**, 30 μm, **e**, 35 μm, and **f**, unstructured QW film.

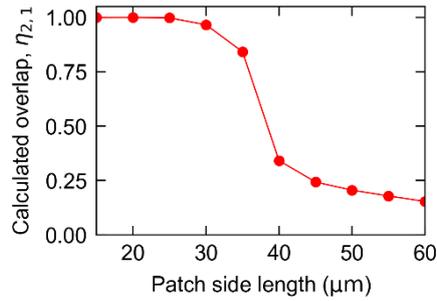

**Figure S6**: Overlap parameter $\eta_{2,1}$ as a function of the side length of the quadratic QW patches. The data are calculated from the simulated near-field distribution of the resonator modes.